\documentclass[a4paper]{article}

\usepackage{INTERSPEECH2021}
\usepackage{lineno,hyperref,multirow,diagbox,multicol,balance,epstopdf,amsmath, subfigure}
\usepackage{url}

\title{Improving Performance of Seen and Unseen Speech Style Transfer \\in End-to-end Neural TTS}
\name{Xiaochun An$^{1}$\thanks{Part of this work has been done during the first author's internship in Microsoft.}, Frank K. Soong$^2$, Lei Xie$^{1\star}$\thanks{$\star$ Corresponding author.}\thanks{This work is supported by the National Key Research and Development Program of China (No. 2020AAA0108600).}}

\address{
  $^1$Audio, Speech and Language Processing Group (ASLP@NPU),\\
School of Computer Science, Northwestern Polytechnical University, Xi'an, China\\
  $^2$Microsoft China}
\email{xiaochunan@npu-aslp.org, frankkps@microsoft.com, lxie@nwpu.edu.cn}

\begin{document}

\maketitle
\begin{abstract}
  End-to-end neural TTS training has shown improved performance in speech style transfer. However, the improvement is still limited by the training data in both target styles and speakers.  Inadequate style transfer performance occurs when the trained TTS tries to transfer the speech to a target style from a new speaker with an unknown, arbitrary style. In this paper, we propose a new approach to style transfer for both seen and unseen styles, with disjoint, multi-style datasets, i.e., datasets of different styles are recorded, each individual style is by one speaker with multiple utterances. To encode the style information, we adopt an inverse autoregressive flow (IAF) structure to improve the variational inference. The whole system is optimized to minimize a weighed sum of four different loss functions: 1) a reconstruction loss to measure the distortions in both source and target reconstructions; 2) an adversarial loss to ``fool" a well-trained discriminator; 3) a style distortion loss to measure the expected style loss after the transfer; 4) a cycle consistency loss to preserve the speaker identity of the source after the transfer. Experiments demonstrate, both objectively and subjectively, the effectiveness of the proposed approach for seen and unseen style transfer tasks. The performance of the new approach is better and more robust than those of four baseline systems of the prior art.
\end{abstract}

\section{Introduction}

Recent advancement of end-to-end neural TTS has demonstrated that it can synthesize very natural, human-like speech~\cite{wang2017tacotron, ping2018deep, shen2018natural, fengyu2019improving, guangzhi2020generating}. The trained neural TTS models usually consist of an encoder-decoder neural network~\cite{sutskever2014sequence, bahdanau2015neural} which can map a text sequence to a sequence of speech frames. Extensions of these models have shown that speech styles (e.g., speaker identity, emotion and prosody), which are essential for expressive and diverse voice generation, can be also modelled and controlled~\cite{Ye2018Transfer, wu2019end-to-end, stanton2018predicting, skerry2018towards, liu2020expressive, lei2021fine}. As applicable scenarios of speech synthesis have rapidly developed, such as the audio reading scenario, there is a growing demands for the single-speaker, multi-style synthesis, where a person can simultaneously speak multiple styles, and yet research in this area is still in its infancy.

Currently most neural TTS systems~\cite{akuzawa2018expressive, xiaochun2019learning, hsu2018hierarchical, habib2020semi-supervised, guangzhi2020fully} are modelled by using a corpus of a single expressive style. Acquiring and annotating a large set of single-speaker speech data with multiple styles for training a neural TTS is usually expensive and time consuming. It is an effective solution to use transfer learning to perform speech style transfer, which allows a speaker to learn the desired style from the data with this style of other speakers without the data of a certain style, and keeps his own timbre consistency. Recently, neural TTS model with global style tokens (GST)~\cite{wang2018style, li2021controllable} or a variational autoencoder (VAE)~\cite{zhang2019learning} has received interests for controlling and transferring speech styles. Theoretically, these models can model any complex styles in a continuous latent space, so that one can control and transfer style by manipulating the latent variables or variational inference from a reference audio.

However, these researches model all speech styles into one style representation, which contains too much interfering information to be robust and interpretable, and lacks the ability to control a specific speech feature independently. When conducting style transfer, one has to transfer all styles whether desired or not, which may not fit the contexts thus hurts generalization. When conducting style control, one can hardly confirm the relationship between the styles and the coefficients of each dimension of the style representations.

Recently, Bian et al.~\cite{bian2019multi} introduce a multi-reference encoder to GST~\cite{wang2018style} and adopt an intercross training scheme, which together ensure that each sub-encoder of the multi-reference encoder independently disentangles and controls a specific style. They show successful style transfer on a multi-style data scenario. However, their intercross training scheme does not guarantee each combination of style classes is seen during training, causing a missed opportunity to learn disentangled representations of styles and sub-optimal results on disjoint, multi-style datasets.

In order to improve style transfer for the combined style that is underrepresented in the dataset, ~\cite{whitehill2020multi} proposes an adversarial cycle consistency training scheme with paired and unpaired triplets to ensure the use of information from all style classes. Unlike intercross training, the scheme sweeps across all combinations of style classes via paired and unpaired triplets. This provides disentanglement of multiple style classes, enabling the model to transfer style in a more faithful manner than existing methods.

Though ~\cite{whitehill2020multi} improves performance of style transfer, it suffers a limitation, similar to ~\cite{bian2019multi}, that can only transfer the style seen during training, and is inadequate to transfer the speech to a target style from a new speaker with an unknown, arbitrary style, thus narrowing down the applicable scenarios of neural TTS systems. In addition, recording training samples for new style (e.g., customer-service style and poetry style) is challenging and labor-intensive, transferring style from one dataset to another (i.e., disjoint, multi-style datasets) is an appealing feature for TTS systems. Therefore, unseen style transfer on disjoint, multi-style datasets needs to be improved.

In this paper, we propose a new approach to style transfer for both seen and unseen styles. As a result, it tackles the single-speaker, multi-style synthesis in a more flexible and convenient manner, and further meets the needs of audio reading scenario. The main contributions of this paper are summarized as follows:

\begin{itemize}
  \item To facilitate seen and unseen style transfer in end-to-end neural TTS, we first adopt an inverse autoregressive flow (IAF) structure~\cite{kingma2016improving} to improve the style representation, and then propose four different loss functions to together make sure the seen and unseen style transfer: 1) using a reconstruction loss to measure the distortions in both source and target reconstructions; 2) injecting an adversarial loss to ``fool" a well-trained discriminator; 3) introducing a style distortion loss to measure the expected style loss after the transfer; 4) incorporating a cycle consistency loss to preserve the speaker identity of the source after the transfer. With the proposed approach, we can transfer the speech to a target style from a new speaker with an unknown, arbitrary style, which does not even need to be seen during training.
  \item The proposed seen and unseen style transfer scheme is used as a data augmentation method to generate a single-speaker, multi-style speech data, which is significant for various speech tasks, such as multi-style TTS and voice conversion.
  \item Our approach outperforms the four prior art baselines and the improvement is confirmed in both subjective and objective tests. The resultant performance of seen and unseen style transfer is better and more robust than the counterpart in the prior art. 
\end{itemize}

\vspace{-13pt}
\section{Proposed approach}
\label{Model architecture}
\begin{figure}[htb]
  \centering
  \includegraphics[width=\linewidth]{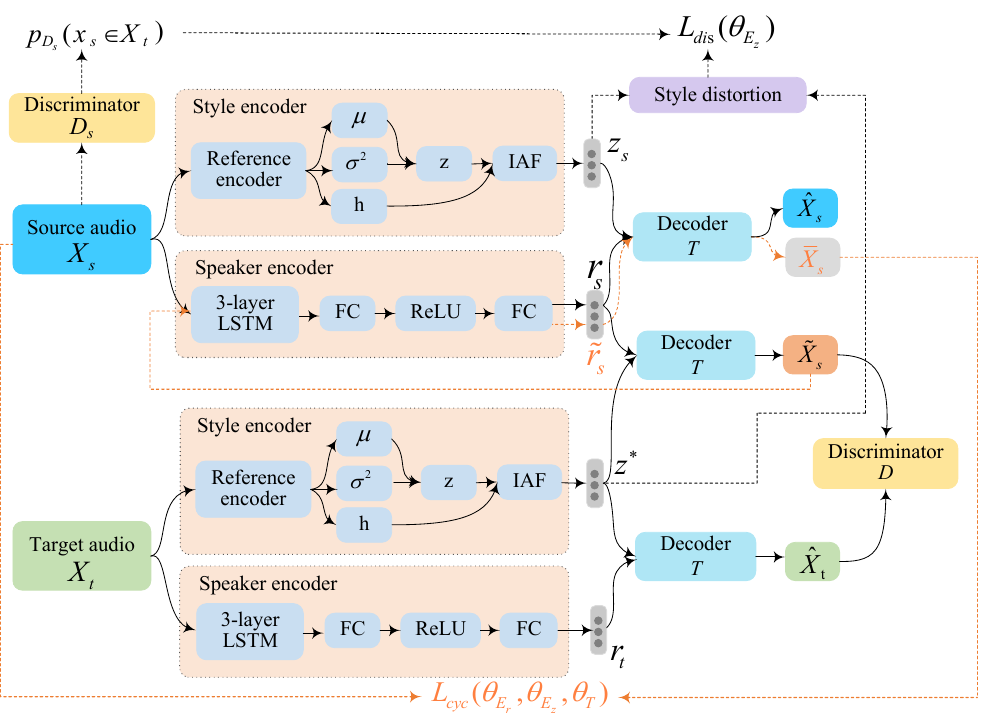}
  \vspace{-15pt}
  \caption{Proposed approach for seen and unseen style transfer.}
  \label{fig1}
  \vspace{-18pt}
\end{figure}

Fig.~\ref{fig1} illustrates our proposed framework to style transfer for both seen and unseen styles, where we adopt Tacotron 2~\cite{shen2018natural} as the decoder and employ the Mel LPCNet vocoder~\cite{Valin2019LPCNET} to reconstruct the waveforms from Mel-spectrogram.
\vspace{-10pt}

\subsection{Preliminary}
\vspace{-2pt}
Our approach dealing with seen and unseen style transfer is built on the encoder-decoder framework. Below we briefly describe the encoder-decoder based style transfer framework.

In the encoder, we assume that each speech utterance $x$ can be decomposed into the style representation, $z \in Z$, and the speaker representation, $r \in R$. Each source utterance, $x_s^{(i)} \in X_s =\{x_s^{(1)} ,\ldots, x_s^{(n)}\}$, has its individual style, $z_s^{(i)}$, while the target utterances, $x_t^{(i)} \in X_t=\{x_t^{(1)} ,\ldots, x_t^{(m)}\}$, share the same style, $z^{\ast}$. We use two encoding functions, $E_z(x)$ and $E_r(x)$, to calculate the $z$ and $r$ of an utterance $x$ respectively: $z_s^{(i)}=E_z(x_s^{(i)})$, $r_s^{(i)}=E_r(x_s^{(i)})$, $z^{\ast}=E_z(x_t^{(j)})$, $r_t^{(j)}=E_r(x_t^{(j)})$.

For the decoder $T$, we leverage Tacotron 2~\cite{shen2018natural} as the decoding function and enforce that for a sample $x_s$, its decoded sequence using $T$ given its speaker representation $r$ and target style representation $z^{\ast}$, should be in the target domain $X_t$. A reconstruction loss is used for maintaining reconstruction fidelity of the utterance in the decoder $T$ as:
\vspace{-4pt}
\begin{equation}
\vspace{-4pt}
\begin{split}
   &\mathcal{L}_{rec}(\theta_{E_r},\theta_{E_z},\theta_{T})\\
   &=\mathbb{E}_{x_s \sim X_s}[-\log p_{T}(x_s |z_s,r_s)]\\
   &+\mathbb{E}_{x_t \sim X_t}[-\log p_{T}(x_t |z^{\ast},r_t)]
  \label{equation1}
\end{split}
\end{equation}
Following GAN~\cite{Ian2014generative, mathieu2016disentangling, guo2019new}, we introduce an adversarial loss to be minimized in decoding and adopt a discriminator $D$, as shown in Fig.~\ref{fig1}, to distinguish between $T(r_s, z^{\ast})$ and $T(r_t, z^{\ast})$, while the task of decoder is to fool the discriminator, as shown in Eq.~\ref{equation2}. The details of each individual component in our model will be described in Section~\ref{Model details}.
\vspace{-4pt}
\begin{equation}
\vspace{-4pt}
\begin{split}
   &\mathcal{L}_{adv}(\theta_{E_r},\theta_{E_z},\theta_{T},\theta_{D})\\
   &=\mathbb{E}_{x_s \sim X_s}[-\log (1-D(T(r_s,z^{\ast})))]\\
   &+\mathbb{E}_{x_t \sim X_t}[-\log D(T(r_t,z^{\ast}))]
  \label{equation2}
\end{split}
\end{equation}

\subsection{Seen and unseen style transfer}

For a sample $x_s \in X_s$, $z_s$ can be an arbitrary value that minimizes the above reconstruction loss and adversarial loss, which may not necessarily capture the utterance style. This will affect the speaker representation, making it not fully represent the speaker identity, which should be invariant with the style.

To address the issue, this paper introduces a style distortion loss into the above framework to constrain that style representation of an utterance should be closer to the target style representation. As shown in Fig.~\ref{fig1}, a discriminator, $D_s$, is first trained to predict whether a given utterance $x$ has the target style with an output probability, $p_{Ds}(x \in X_t)$. When learning the style representation $z_s$, we then enforce that the distortion between this style representation $z_s$ and target style representation $z^{\ast}$ should be consistent with the output probability of $D_s$. Here, we use the $L_2$ norm to measure the style distortion, $d(z_s, z^{\ast})=\|z_s -z^{\ast}\|_2$, and want to have the style distortion positively correlated with $1-p_{Ds}(x_s \in X_t)$. To incorporate this idea into our model, we adopt a probability density function, which is modelled with a standard normal distribution in our experiments, to evaluate the style distortion loss. Intuitively, when an utterance $x_s$ have a large output probability $p_{Ds}(x_s \in X_t)$, our model encourages a large probability density and a small style distortion. This means $z_s$ will be closer to $z^{\ast}$, and the style distortion loss can finally be written as:
\vspace{-5pt}
\begin{equation}
\vspace{-5pt}
\begin{split}
   &\mathcal{L}_{dis}(\theta_{E_z})=\mathbb{E}_{x_s \sim X_s}[p_{Ds}(x_s \in X_t)d(z_s, z^{\ast})^2]
  \label{equation3}
\end{split}
\end{equation}
where $D_s$ is a pre-trained model trained with a portion of the training data. Here, if we integrate $D_s$ into our training, we may start with a $D_s$ with a low accuracy, and then our model is inclined to optimize a wrong style distortion loss for many epochs and gets stuck into a poor local optimum.

In addition, the discriminator $D_s$ can only constrain the generated utterance to be aligned with the target style, but cannot guarantee to keep the speaker of the source utterance intact. To address the problem, we further introduce a cycle consistency loss~\cite{zhu2017unpaired, Liumeng2021on} to our model shown in Fig.~\ref{fig1}, which requires that a transferred utterance should preserve the speaker identity of its source utterance, and thus it is enable to recover the source utterance in a cyclic manner. The cycle consistency loss is shown as follows:
\vspace{-4pt}
\begin{equation}
\vspace{-4pt}
\begin{split}
   &\mathcal{L}_{cyc}(\theta_{E_r},\theta_{E_z},\theta_{T})\\
   &=\mathbb{E}_{x_s \sim X_s}[-\log p_{T}(x_s| E_r (\widetilde{x}_s), z_s)]\\
   &+\mathbb{E}_{x_t \sim X_t}[-\log p_{T}(x_t| E_r (\widetilde{x}_t), z^{\ast})]
  \label{equation4}
\end{split}
\end{equation}
where $\widetilde{x}_s$ is the transferred utterance from a source sample $x_s$ and has the target style $z^{\ast}$. We encode the $\widetilde{x}_s$ with the speaker encoder $E_r(\widetilde{x}_s)$ to obtain its speaker representation $\widetilde{r}_s$, which is combined with its source style $z_s$ for decoding. Here, we expect that the source utterance can be generated with a high probability. For a target sample $x_t$, although we do not aim to change its style in our model, similar to $x_s$, we still calculate its cycle consistency loss for the purpose of additional regularization and hope that the target utterance $x_t$ should be generated. To summarize, the final form of our loss function is:
\vspace{-4pt}
\begin{equation}
\vspace{-4pt}
\begin{split}
   &\mathcal{L}(\theta_{E_r},\theta_{E_z},\theta_{T},\theta_{D})\\
   &=\alpha\mathcal{L}_{rec}+\beta\mathcal{L}_{adv}+\gamma\mathcal{L}_{dis}+\lambda\mathcal{L}_{cyc}
  \label{equation5}
\end{split}
\end{equation}
where $\alpha=\beta=\lambda=1.0$ and $\gamma=5.0$ are preset weights for balancing the different loss terms. In our experiments, we find the results are insensitive to these parameters through cross-validation. 

\vspace{-8pt}
\section{Experiments}
\label{experiments}

This paper focuses on disjoint, multi-style datasets and an internal Chinese corpus is used in experiments: source data contains examples of four styles (i.\,e., reading style, broadcasting style, talking style and story style) from four different speakers whereas target data contains samples of two styles (i.\,e., customer-service style and poetry style) from two different speakers. This represents a minimalistic scenario of the disjoint, multi-style datasets: a single model must be able to properly transfer an arbitrary and unknown style to target style where there is no variation of speaker identity. The corpus contains 20,698 samples ($\sim$ 23.3 hours) and each style contains 4,000 samples except for poetry style. There are 698 samples in poetry style. We remove long silence ($\textgreater$ 0.1 sec) at the beginning and ending of each utterance. Mel spectrum is extracted as target speech representations with a Hanning window of 50 ms and 12.5 ms frame shift. Phoneme sequences are used as the text input. For all different systems in our experiments, we train $\sim$ 220k steps with a single Nvidia Tesla P40 GPU. The models on which we conduct experiments include:
\begin{itemize}
  \item GST: similar to ~\cite{wang2018style}, we introduce ``global style tokens" (GSTs) to Tacotron 2~\cite{shen2018natural} to perform various style control and transfer tasks, and make a fair comparison.
  \item VAE: we incorporate VAE~\cite{zhang2019learning} into Tacotron 2~\cite{shen2018natural} to learn the latent representation of speaking styles to guide the style in synthesizing speech.
  \item MRF-IT: we augment a multi-reference encoder structure into GST-Tacotron 2~\cite{bian2019multi} and adopt intercross training approach to control and transfer desired speech styles.
  \item MRF-ACC: we adopt an adversarial cycle consistency training scheme for multi-reference neural TTS stylization~\cite{whitehill2020multi} to ensure the use of information from all style classes and achieve style transfer.
  \item Proposed model: we adopt an IAF structure~\cite{kingma2016improving} to improve style representation, and introduce four different loss functions, including: reconstruction loss, adversarial loss, style distortion loss and cycle consistency loss, to together make sure the seen and unseen style transfer on disjoint, multi-style datasets.
\end{itemize}

We conduct mean opinion score (MOS) and preference listening tests (ABX) of speech quality to evaluate the reconstruction performance of different experimental systems. An ABX test of style similarity is conducted to assess the style conversion performance, where subjects are asked to choose the speech samples which sound closer to the target style in terms of style expression. We further conduct a comparative mean opinion score (CMOS) test of speaker similarity to evaluate how well the transferred speech matches that of the source speaker. For each system, we randomly select the reading style as the seen style and make two experiments: from reading style to customer-service style (R2C) and reading style to poetry style (R2P). We randomly choose an unique Taiwanese-reading style from a new female speaker as the unseen style, and conduct two tests from Taiwanese-reading style to customer-service style (TR2C) and Taiwanese-reading style to poetry style (TR2P) to assess the performance of unseen style transfer.
\vspace{-6pt}
\subsection{Model details}
\label{Model details}
\vspace{-2pt}
The style encoder contains a reference encoder and an IAF~\cite{kingma2016improving} flow to extract underlying stylistic properties, which form a more discriminative and expressive latent style representation. Similar to ~\cite{zhang2019learning}, the reference encoder consists of a stack of six 2-D convolutional layers cascaded with one unidirectional 128-unit GRU layer, and the architecture of the IAF is the same as that in ~\cite{Aggarwal2020using}. The output of reference encoder network is then used to estimate initial mean $\mu$, initial variance $\delta$, and hidden output $h$. Afterward, the initial latent variable $z$ along with hidden output $h$ is provided to $k$ steps of IAF transformation to obtain flexible posterior probability distribution with latent variable $z_s$. For the speaker encoder, we use a 3-layer LSTM with the projection operations shown in Fig.~\ref{fig1}. In our model, we adopt Tacotron 2~\cite{shen2018natural} as the decoder, which takes the concatenation of the speaker and style representations as the initial hidden state. As for the discriminator $D$, we follow the architecture of the discriminator in ~\cite{guo2019new}. The pre-trained discriminator $D_s$ used in the style distortion loss has the same structure as the style encoder followed by a sigmoid output layer.
\vspace{-7pt}
\subsection{Experimental results}
\begin{figure}[t]
\vspace{-6pt}
  \centering
  \includegraphics[width=\linewidth]{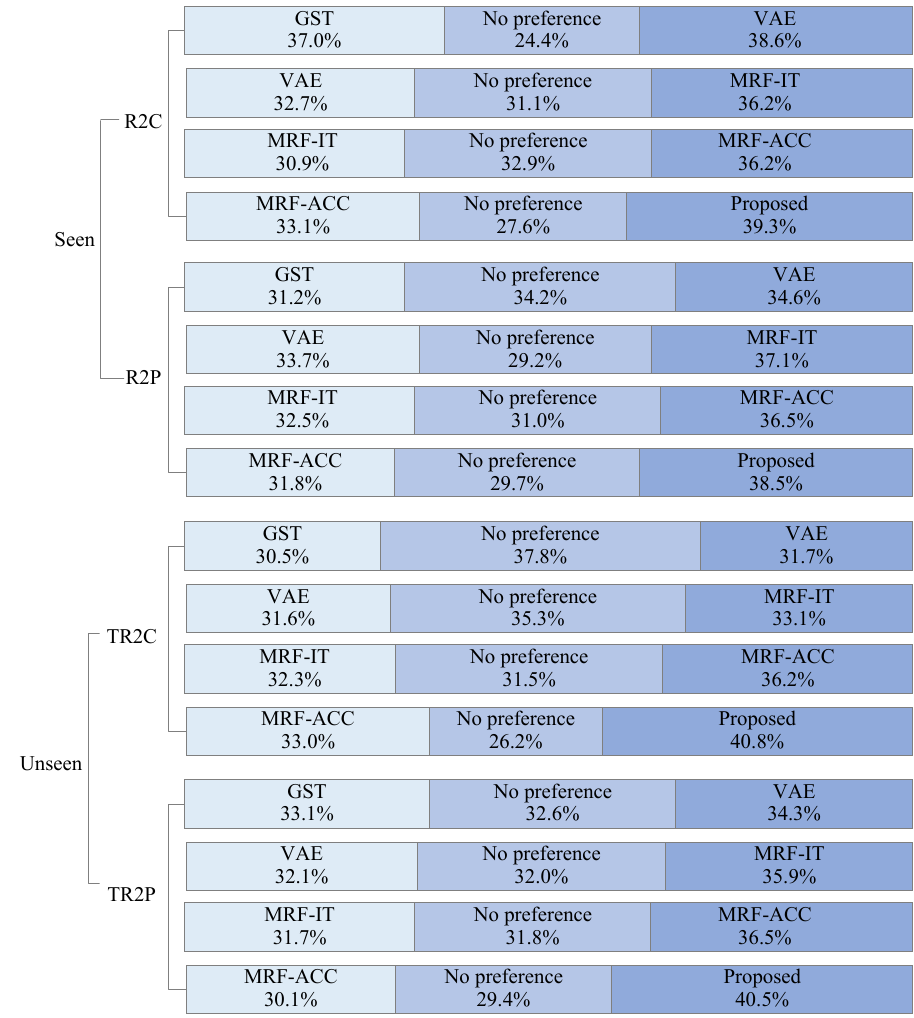}
  \vspace{-14pt}
  \caption{ABX preference results for speech quality.}
  \vspace{-17pt}
  \label{fig2}
\end{figure}

\vspace{-2pt}
\textbf{Speech quality} We use the seen and unseen test sets, which contain 20 sentences of each style, respectively, to compare the performance of all models in speech quality and naturalness with the MOS and ABX listening tests\footnote{Samples can be found at \url{https://xiaochunan.github.io/transfer/index.html}}. The 15 subjects need to mark a sentence unintelligible when any part of it is unintelligible in listening. Table~\ref{table1} and Fig.~\ref{fig2} show the results of these two subjective evaluations. The proposed model outperforms the baseline models on both seen and unseen style transfer tasks. The performance of the proposed model on unseen style transfer is much better than other models. The results show a better generalization of the proposed model on the unseen style transfer. These observations validate the effectiveness of our proposed model in terms of speech quality. Seen style transfer performs better speech quality than the unseen style transfer. The difference is partially due to the smaller speech data set of the poetry style, hence its MOS score is lower than that of the customer-service style.

\vspace{-0.5pt}
\begin{table}[!t]
\vspace{-20pt}
  \caption{MOS results with 95 \% confidence interval of seen and unseen style transfer from different models.}
  \vspace{-6pt}
  \label{table1}
  \centering
  \scalebox{0.87}{
  \begin{tabular}{lllll}
    \toprule
    \multirow{2}{*}{Models} &\multicolumn{2}{l}{Seen style transfer} &\multicolumn{2}{l}{Unseen style transfer}\\
    \cline{2-5}
             &R2C       &R2P      &TR2C     &TR2P\\
    \midrule
    GST      &3.33$\pm$0.06   &3.27$\pm$0.08   &2.91$\pm$0.10   &2.82$\pm$0.06~~~             \\
    VAE      &3.42$\pm$0.04   &3.38$\pm$0.03   &2.95$\pm$0.05   &2.87$\pm$0.05~~~             \\
    MRF-IT   &3.54$\pm$0.07   &3.46$\pm$0.07   &3.07$\pm$0.06   &2.98$\pm$0.12~~~             \\
    MRF-ACC   &3.75$\pm$0.11   &3.69$\pm$0.09   &3.57$\pm$0.04   &3.48$\pm$0.07~~~             \\
    Proposed &\textbf{3.96$\pm$0.03}   &\textbf{3.89$\pm$0.12}   &\textbf{3.80$\pm$0.05}   &\textbf{3.73$\pm$0.02}~~~             \\
    \bottomrule
  \end{tabular}
  }
\end{table}

\begin{figure}[t]
\vspace{-20pt}
  \centering
  \includegraphics[width=\linewidth]{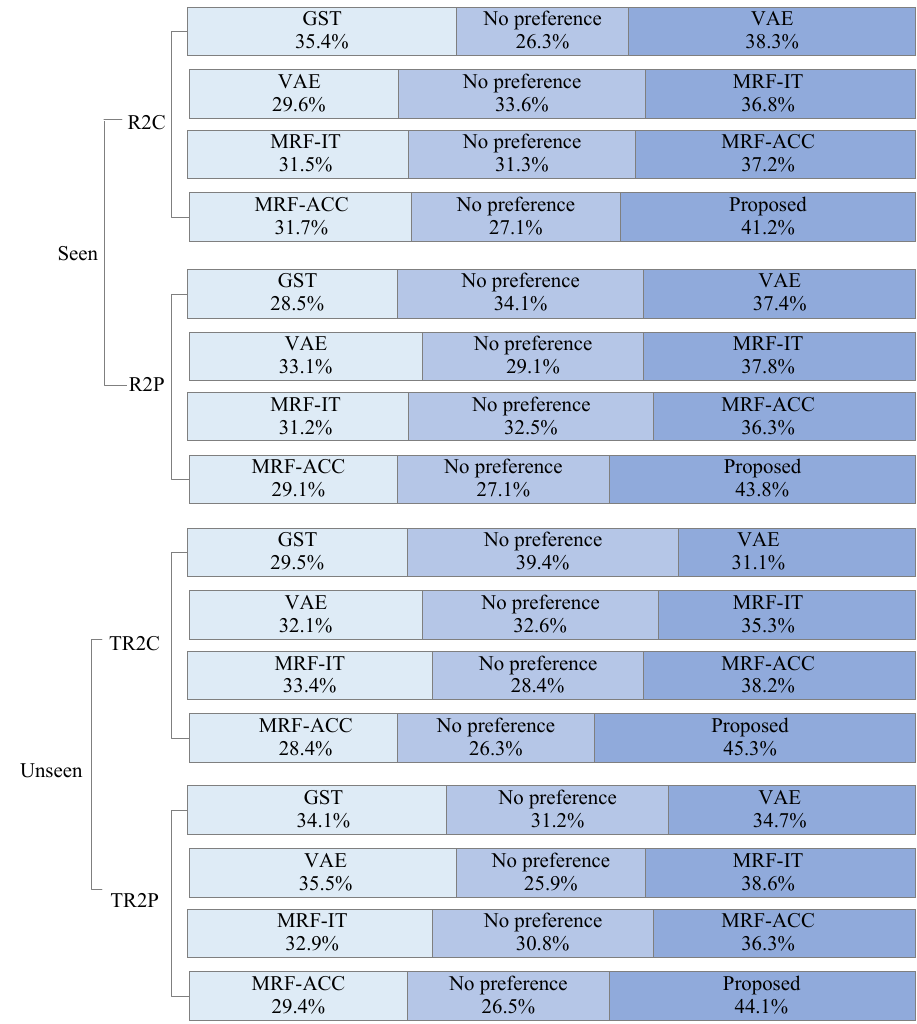}
  \vspace{-14pt}
  \caption{ABX preference results for style similarity.}
  \vspace{-15pt}
  \label{fig3}
\end{figure}

\textbf{Style similarity} Adopting the same test sets, we conduct an ABX test of style similarity to assess the style conversion performance. The same 15 listeners are asked to choose which speech sample sounds closer to the target style. The results are shown in Fig.~\ref{fig3}, where the listeners give preference to the proposed system, showing the proposed method improves performance of style transfer. For the unseen style, we find that GST, VAE and MRF-IT models, in most cases, fail to transfer unseen style of the Taiwanese-reading style to the target style of customer service or poetry style. The MRF-ACC system, the best of all baseline system, is still significantly inferior to the proposed model in the style similarity test. The results demonstrate the effectiveness of our proposed approach for both seen and unseen style transfer.

\textbf{Speaker similarity} To evaluate how well the transferred speech matches that of the source speaker's timbre, we conduct CMOS tests between the proposed model and each baseline by using the same test sets. The same 15 listeners are asked to select audio that represents a closer speaker to source audio. Table~\ref{table2} reports CMOS results for speaker similarity, where score of the proposed model is fixed to 0. Comparisons between the proposed model and the baseline models show that our approach delivers better speaker similarity performance than all baseline models, on both seen and unseen style transfer tasks. We further adopt the cosine distance to calculate the similarity between the speaker embedding of a transferred sample and the speaker embedding of a randomly selected ground truth utterance from the same speaker to objectively measure the speaker conversion performance. The results are shown in Table~\ref{table3} where we observe that the proposed model delivers a higher similarity than the other models. On the unseen style transfer of TR2C and TR2P, the GST, VAE and MRF-IT, MRF-ACC models are not capable of keeping the speaker's timbre, resulting in low similarity scores. The best baseline, MRF-ACC system, still performs significantly worse than the proposed model.

\begin{table}[htb]
\vspace{-6pt}
  \caption{CMOS results for speaker similarity.}
  \vspace{-6pt}
  \label{table2}
  \centering
  \scalebox{0.87}{
  \begin{tabular}{lllll}
    \toprule
    \multirow{2}{*}{Models} &\multicolumn{2}{l}{Seen style transfer} &\multicolumn{2}{l}{Unseen style transfer}\\
    \cline{2-5}
             &R2C              &R2P             &TR2C            &TR2P\\
    \midrule
    Proposed &0        &0       &0       &0~~~             \\
    GST      &-0.78    &-0.80   &-0.90   &-0.92~~~         \\
    VAE      &-0.55    &-0.56   &-0.65   &-0.68~~~             \\
    MRF-IT   &-0.34    &-0.35   &-0.40   &-0.42~~~             \\
    MRF-ACC  &-0.16    &-0.20   &-0.26   &-0.28~~~             \\
    \bottomrule
  \end{tabular}
  }
\end{table}

\begin{table}[htb]
\vspace{-12pt}
  \caption{Cosine distance results for speaker similarity.}
  \vspace{-6pt}
  \label{table3}
  \centering
  \scalebox{0.87}{
  \begin{tabular}{lllll}
    \toprule
    \multirow{2}{*}{Models} &\multicolumn{2}{l}{Seen style transfer} &\multicolumn{2}{l}{Unseen style transfer}\\
    \cline{2-5}
             &R2C              &R2P             &TR2C            &TR2P\\
    \midrule
    GST      &0.29    &0.28   &0.19   &0.17~~~             \\
    VAE      &0.35    &0.34   &0.21   &0.20~~~             \\
    MRF-IT   &0.45    &0.43   &0.34   &0.32~~~             \\
    MRF-ACC  &0.57    &0.56   &0.42   &0.40~~~             \\
    Proposed &\textbf{0.69}    &\textbf{0.68}   &\textbf{0.65}   &\textbf{0.64}~~~             \\
    \bottomrule
  \end{tabular}
  }
\end{table}

\vspace{-14pt}
\section{Conclusions}
\label{conclusion}
This paper investigates how to train an end-to-end neural TTS for both seen and unseen speech style transfer with disjoint training datasets. We adopt an IAF structure to encode the style information and optimize a weighed sum of four, reconstruction, adversarial, style and cycle consistency, loss functions. The weighted total loss is minimized to optimize style transfer performance. The source speaker's identity or the voice timbre is well preserved by the additional cycle consistency loss. Experiments demonstrate that the proposed approach, for both seen and unseen style transfer, can outperform four other systems of the prior art, both objectively and subjectively.

\bibliographystyle{IEEEtran}

\bibliography{mybib}


\end{document}